\documentclass[aps,amssymb,pra,twocolumn,showpacs,floatfix]{revtex4}

\usepackage{epsf}
\usepackage{amsmath}
\usepackage{graphicx}
\begin{document}
         
\title {On formation of long-living states}

\author{Boris I. Ivlev}

\affiliation
{Instituto de F\'{\i}sica, Universidad Aut\'onoma de San Luis Potos\'{\i},
San Luis Potos\'{\i}, 78000 Mexico\\}

\begin{abstract}
The motion of a particle in the potential well is studied when the particle is attached to the infinite elastic string. This is generic with the
problem of dissipative quantum mechanics investigated by Caldeira and Leggett \cite{LEGG}. Besides the dissipative motion there is another scenario 
of interaction of the string with the particle attached. Stationary particle-string states exist with string deformations accompanying the particle.
This is like polaronic states in solids. Our polaronic states in the well are non-decaying and with continuous energy spectrum. Perhaps these 
states have a link to quantum electrodynamics. Quantum mechanical wave function, singular on some line, is smeared out by electron ``vibrations'' 
due to the interaction with photons. In those anomalous state the smeared singularity position would be analogous to the place where the particle 
is attached to the string. 

\end{abstract} \vskip 1.0cm

\pacs{03.65.-w, 03.65.Ge}

\maketitle

\section{Introduction}
\label{intr}
Discrete energy levels of the electron in a potential well become of a finite width under the interaction with photons. This provides finite 
lifetimes of levels with respect to photon emission. Photons are emitted until the electron reaches the ground state level. This level has zero 
width. It is slightly shifted (the Lamb shift) due to the electron-photon interaction \cite{LANDAU3}. 

Discrete levels can be of different lifetimes. When the certain level is long-living this results in possibility of laser generation. Broadening
of the energy level is characterized by the imaginary part of its energy. The level widths in various atoms are known \cite{KRAUSE}. 

There is a question which does not seem conventional. Are there some conditions to get long-living states with the continuous energy spectrum in 
a potential well? Such states would provide unusual laser generation. This phenomenon may occur solely due to interaction with a reservoir since 
in quantum mechanics levels in a well are discrete. 

As shown in the paper, the answer to that question is positive. The motion of a particle in the well is studied when the particle is attached
to the infinite elastic string. The moving particle causes elastic waves of the string which carry away the energy. The energy dissipation 
provides friction motion of the particle. This problem is generic with one in paper \cite{LEGG} studying dissipative quantum mechanics. See also
\cite{LAR1,SCHMID,MELN,CHAKR,HANG1,LEGG1,KOR,IVLEV1,HANG2,KAGAN,LAR2,WEISS}.

Besides such dissipative motion there is another scenario of interaction of the string with the attached particle. That regime is not dissipative.
The joint particle-string state is stationary and string deformations accompany the particle. This reminds polaronic state in a solid when the 
electron is ``dressed'' by phonons \cite{KITT}. Our polaronic states in the well are continuously distributed in energy which does not have an
imaginary part. This means that polaronic states of all their energies are non-decaying. 

One can qualitatively explain why photons (string sound waves) are not emitted by polaronic states. This is due to the ``hard'' connection of the 
particle to the string. Emission of waves would result in oscillations of the string including the point of particle attachment. This increases the 
particle kinetic energy preventing it to lose its total energy and therefore resulting in non-decaying states.

The issue is that nature allows the continuous non-decaying energy spectrum in a potential well. This conclusion is based on the exact solution of 
the particle-string problem obtained in this paper. There is no contradiction to quantum mechanics since the particle is coupled to the reservoir. 

The solution obtained would be just a model but perhaps there is a link from those polaronic states to quantum electrodynamics. 

In quantum electrodynamics radiative corrections are small \cite{LANDAU3} and one can say that the electron is not connected ``hard'' to 
electromagnetic coordinates as in the string case. Therefore shifts of discrete energy levels of the electron (the Lamb shift \cite{LANDAU3}) are
small. This is accompanied by a small broadening of higher levels resulting in photons emission. The Lamb shift occurs since, due to the photon 
influence, the electron ``vibrates'' within the narrow region of $10^{-11}$cm \cite{MIGDAL,BOY}. In that way it probes various parts of the 
potential and therefore slightly changes its energy. Thus in quantum electrodynamics polaronic states are impossible at first sight. 

However, in quantum electrodynamics there are the certain conditions when states of the polaronic type do not seem definitely excluded. The 
electron wave function in quantum mechanics can have a formal singularity on some line. Such state is not physical. But under electron 
``vibrations'' the singularity is smeared out within the above narrow region. This anomalous electron-photon state (of the polaronic type), if it
exists, is physical. 

That narrow region would play a role of a point where the electron is connected ``hard'' to electromagnetic coordinates and is dragged by them. 
One can treat the electron to be localized in that region. Under photon emission the narrow region would oscillate increasing the electron kinetic 
energy. As in the string case this prevents the electron to lose its total energy resulting in non-decaying states.

Therefore in quantum electrodynamics one can put a question on possibility of non-decaying states of the electron in a potential well. These states 
are of continuous energy spectrum since the condition of absence of singularities, leading to levels quantization, is not imposed.

In Sec.~\ref{ela} the exact solution for the system particle-string is obtained. In Sec.~\ref{disc} the hypothesis of the link to quantum 
electrodynamics is discussed.
\begin{figure}
\includegraphics[width=4.5cm]{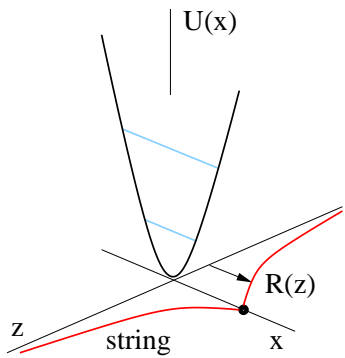}
\caption{\label{fig1}Particle can move along the x-direction only. This particle, acted by the potential $U(x)$, is attached to the elastic string
with displacements $R(z)$. The particle displacement is $x=R(0)$. Exact states of the system are localized. In the absence of the string two 
discrete levels in the well are shown.}
\end{figure}
\section{PARTICLE ON A STRING}
\label{ela}
Below we study the elastic string placed along the $z$ axis. Transverse displacements of the string are $R(z,t)$ as shown in Fig.~\ref{fig1}.  
The particle of the mass $m$ is attached to the string at the point $z=0$ and moves together with the string. The particle coordinate is 
$x(t)=R(0,t)$. The potential energy $U(x)$ depends on the particle coordinate only. 

The energy of the string with the particle has the form
\begin{eqnarray}
\nonumber
&&E=\int dz\bigg\{\frac{1}{2}\left[\rho+m\delta(z)\right]\dot{R}^2+
\frac{\rho s^2}{2}\left(\frac{\partial R}{\partial z}\right)^2\\
\label{1}
&&~~~~~+\delta(z)U(R)\bigg\}.
\end{eqnarray}
In Eq.~(\ref{1}) $\rho$ is the mass density of the string and $s$ is the velocity of elastic waves. The second term is the elastic energy. The 
absence of the string corresponds to the limit $\rho=0$ and then the energy takes its usual form
\begin{equation}
\label{2}
E= \frac{m}{2}\,\dot{x}^2+U(x).
\end{equation}
\subsection{Classical limit} 
In the classical limit the function $R(z,t)$ is a macroscopic variable. It obeys the equation obtained by the variation of the functional (\ref{1})
\begin{equation}
\label{100}
\left[\rho+m\delta(z)\right]\ddot{R}-\rho s^2\frac{\partial^2R}{\partial z^2}+U'(R)\delta(z)=0.
\end{equation}
The integration of the singular part in Eq.~(\ref{100}) results in 
\begin{equation}
\label{101}
m\ddot{x}+U'(x)-\rho s^2\left(\frac{\partial R}{\partial z}\bigg|_{z=+0}-\frac{\partial R}{\partial z}\bigg|_{z=-0}\right)=0.
\end{equation}
At $z\neq 0$ the wave equation holds
\begin{equation}
\label{102}
\ddot{R}-s^2\frac{\partial^2 R}{\partial z^2}=0
\end{equation}
\begin{figure}
\includegraphics[width=6cm]{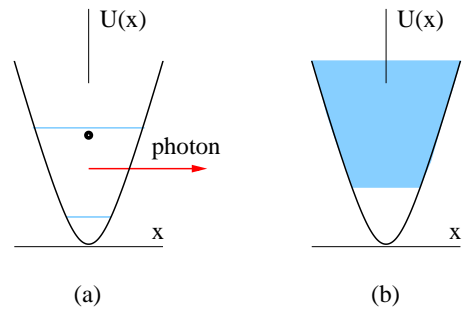}
\caption{\label{fig2}(a) Dissipative motion of the particle in the well. The electron-photon interaction provides a finite width of the upper level 
resulting in the photon emission. (b) Polaronic states. Energy spectrum of the particle, attached to the string, is continuous and non-decaying 
(zero imaginary part of energy).}
\end{figure}
with the solution 
\begin{equation}
\label{103}
R(z,t)=x\left(t-\frac{|z|}{s}\right).
\end{equation}
This solution, in the form of two waves running from the point $z=0$, corresponds to the condition $R(0,t)=x(t)$. The dynamic equation for $x$
gets the form
\begin{equation}
\label{104}
m\ddot{x}+U'(x)+\eta\dot{x}=0,
\end{equation}
where $\eta=2\rho s$. Eq.~(\ref{104}) corresponds to the particle with friction provided by the radiation mechanism. The energy of the moving
particle is carried away by sound waves from the point $z=0$. 
\subsection{Discrete coordinates} 
It is convenient to consider the string of the finite length $L$. In the finale results the limit $L\rightarrow\infty$ is taken. According to this,
one can use the Fourier series
\begin{equation}
\label{3}
R(z,t)=\frac{1}{L}\sum_{n}R_n(t)\exp\left(\frac{2\pi in}{L}z\right).
\end{equation}
In the representation (\ref{3}) the energy (\ref{1}) takes the form
\begin{equation}
\label{4}
E=\frac{\rho}{2L}\sum_n\left(\dot{R}^{2}_{n}+\omega^{2}_{n}R^{2}_{n}\right)+
\frac{m}{2}\,\dot{x}^2+U(x),
\end{equation}
where $\omega_n=2\pi |n|s/L$. In Eq.~(\ref{4}) the $R$-part is not separated from the $x$-part due to the connection condition following from 
(\ref{3})
\begin{equation}
\label{5}
x(t)=\frac{1}{L}\sum_{n}R_n(t).
\end{equation}
It is convenient to separate the total $R_n$ by two parts
\begin{equation}
\label{6}
R_n=\frac{r_n+is_n}{\sqrt{2}},
\end{equation}
where $r_{-n}=r_n$ and $s_{-n}=-s_n$. Then the energy (\ref{4}) reads
\begin{eqnarray}
\nonumber
&&E=\frac{\rho}{2L}\sum^{\infty}_{n=1}\left(\dot{s}^{2}_{n}+\omega^{2}_{n}s^{2}_{n}\right)+
\frac{m}{2}\,\dot{x}^2+U(x)\\
\label{7}
&&~~~+\frac{\rho}{2L}\left\{\sum^{\infty}_{n=1}\left(\dot{r}^{2}_{n}+\omega^{2}_{n}r^{2}_{n}\right)
+\frac{1}{2}\,\dot{r}^{2}_{0}\right\}.
\end{eqnarray}
As follows from (\ref{5}), 
\begin{equation}
\label{8}
r_0=xL\sqrt{2}-2\sum^{\infty}_{n=1}r_n\,.
\end{equation}
We see that the $x$-mode is connected to $r$-mode only. The $s$-mode is independent and corresponds to usual string waves. For this reason it
can be omitted. 
\subsection{Dissipative quantum mechanics}
In multi-dimensional quantum mechanics for variables $x$ and $r_n$ ($n=1,2,...$) one can subsequently integrate out the variables $r_n$ 
by simple Gaussian integration \cite{LEGG}. As a result, the effective action in terms of $x$ will get the known Caldeira-Leggett form \cite{LEGG}.
That action allows to describe dissipative quantum systems, that is interacting with reservoir. The classical dissipative limit corresponds 
to Eq.~(\ref{104}). This problem is well investigated in literature \cite{LAR1,SCHMID,MELN,CHAKR,HANG1,LEGG1,KOR,IVLEV1,HANG2,KAGAN,LAR2,WEISS} 
and we do not repeat here known results. Instead one can focus on another aspect of interaction with a reservoir.  
\subsection{Particle energy}
Below we consider the potential in the form of harmonic oscillator $U(x)=m\Omega^2x^2/2$. The total energy (\ref{7}) goes over into two quadratic 
forms with respect to coordinates and their time derivatives. After diagonalization of the forms the energy (\ref{7}) is
\begin{equation}
\label{9}
E=\frac{\rho}{2L}\sum_{q}(\dot{\eta}^{2}_{q}+\omega^{2}_{q}\eta^{2}_{q})\,,
\end{equation}
where $\eta_q$ are the certain variables as in Appendix. The frequency $\omega_q$ depends on the new wave vector $q$ according to the relation
(see Appendix)
\begin{equation}
\label{10}
\Omega^2-\omega^{2}_{q}=\frac{\eta\omega_q}{m}\tan\left(\frac{L\omega_q}{2s}\right).
\end{equation}
The quantization of the new wave vector follows from (\ref{10}) 
\begin{equation}
\label{11}
q_n=\frac{2\pi n}{L}-\frac{2}{L}\arctan\left(\frac{msq_n}{\eta}-\frac{m\Omega^2}{\eta sq_n}\right).
\end{equation}

Formally calculating $\partial n/\partial q_n$ one can establish the connection
\begin{equation}
\label{12}
\sum_nF(\omega_q)=\int^{\infty}_{0}d\omega\,\nu(\omega)F(\omega),
\end{equation}
where the frequency distribution is
\begin{equation}
\label{13}
\nu(\omega)=\nu_0+\frac{\eta}{\pi}\,\frac{m\omega^2+m\Omega^2}{(m\omega^2-m\Omega^2)^2+\eta^2\omega^2}\,.
\end{equation}
Here $\nu_0=L/(2\pi s)$ is referred to the string without the particle. 

The energy (\ref{9}) consists of contributions of independent oscillators. Suppose all frequencies to be separated by groups $\{\omega\}_1$, 
$\{\omega\}_2$, ... Oscillators with frequencies of the group $l$ are in the $l$th excited state. The total energy can be separated by two parts 
$E=E_0+E_p$ where 
\begin{equation}
\label{13a}
E_p=\int^{\infty}_{0}\hbar\omega\left(\frac{1}{2}+\sum^{\infty}_{l=1}N_{l\omega}\right)(\nu-\nu_0)(\omega) d\omega
\end{equation}
is the particle contribution. Each $N_{l\omega}\leq l$ accounts for the frequency distribution in the group $\{\omega\}_l$. When, for example, all 
oscillators are in the first excited state then only $N_{1\omega}$ is not zero and $N_{1\omega}=1$ for all frequencies. Each separation on frequency 
groups corresponds to the certain energy (\ref{13a}). The energy spectrum is continuous.

The string energy, without the particle, $E_0$ is given by analogous equation but with the frequency distribution $\nu_0$ related to the free 
string. The energy $E_0\sim L$ is strongly divergent at large frequencies. 

In the limit $\eta\ll m\Omega$ of small particle dissipation the particle energy $E^{(0)}_{p}$ of the ground state ($N_{l\omega}=0$), according to 
(\ref{13a}), is
\begin{equation}
\label{13b}
E^{(0)}_{p}=\frac{\hbar\Omega}{2}+\frac{\hbar\eta}{2\pi m}\ln\frac{\omega_{max}}{\Omega}\,.
\end{equation}
Here the second term is due to connection to the string. It diverges logarithmically and should be cut off by the large frequency $\omega_{max}$. 
This frequency can be related to the discreteness $a\sim 1/q_{max}$ of the string $\omega_{max}\sim sq_{max}$. 

Correction to the quantum mechanical ground state energy in Eq.~(\ref{13b}) reminds the Lamb shift in hydrogen atom \cite{LANDAU3,MIGDAL}. In that 
case the shift is also proportional to the similar logarithm with $mc^2/\hbar$ instead of $\omega_{max}$.
\subsection{Fluctuation dissipation theorem}
As follows from Eq.~(\ref{5}), 
\begin{equation}
\label{14}
x=\frac{\sqrt{2}}{L}\sum_qu_{0q}\eta_{q}\,.
\end{equation}
The mean squared value of $x$, after the average on fluctuations, is
\begin{equation}
\label{15}
\langle x^2\rangle= \frac{2}{L^2}\sum_q\frac{\eta^2\omega^{2}_{q}}{(m\omega^{2}_{q}-m\Omega^2)^2+\eta^2\omega^{2}_{q}}\,\langle\eta^{2}_{q}\rangle.
\end{equation}
Besides quantum fluctuations one can consider also thermal ones. It follows from (\ref{9}) that 
\begin{equation}
\label{16}
\langle\eta^{2}_{q}\rangle=\frac{L\hbar}{2\rho\omega_q}\cot\frac{\hbar\omega_q}{2T}
\end{equation}
as for a harmonic oscillator. One should use the density of states $\nu_0$. The mean squared displacement (\ref{15}) corresponds to the usual
fluctuation dissipation theorem
\begin{equation}
\label{17}
\langle x^2\rangle=\frac{i\hbar}{2\pi}\int^{\infty}_{-\infty}\cot\frac{\hbar\omega}{2T}\,
\frac{d\omega}{m\omega^{2}-m\Omega^2+i\eta\omega}\,.
\end{equation}
\subsection{Distribution of string displacements}
In Fig.~\ref{fig1} the instant string deformation is schematically shown. The mean displacements of the particle and string coordinates are zero.
There is a reason to look at the mean squared displacements. 

As follows from Eqs.~(\ref{3}), (\ref{5}), and (\ref{8}), the string displacement is
\begin{equation}
\label{105}
R(z)=x+\frac{\sqrt{2}}{L}\sum^{\infty}_{n=1}r_n\left[\cos\left(\frac{2\pi nz}{L}\right)-1\right].
\end{equation}
With the use of Eqs.~(\ref{A2}) and (\ref{14}) it follows
\begin{equation}
\label{106}
R(z)=\frac{\sqrt{2}}{L}\sum_q\eta_q\sum^{\infty}_{n=1}u_{nq}\left[\cos\left(\frac{2\pi nz}{L}\right)-1\right].
\end{equation}
Accounting for Eqs.~(\ref{A11}) and (\ref{A10}) one can obtain
\begin{equation}
\label{107}
R(z)=\frac{\sqrt{2}}{L}\sum_q\eta_qu_{0q}\left[\cos\frac{z\omega_q}{s}+
\frac{m(\Omega^2-\omega^{2}_{q})}{\eta\omega_q}\sin\frac{|z|\omega_q}{s}\right]
\end{equation}
The mean squared displacement of the string is
\begin{eqnarray}
\nonumber
\langle R^2(z)\rangle=\frac{2\hbar}{\pi\eta}\int^{\infty}_{0}\frac{d\omega_q}{\omega_q}\left(\frac{\rho\omega_q}{\hbar L}\,\langle\eta^{2}_{q}
\rangle\right)\Bigg\{\sin^2\frac{z\omega_q}{s}\\
\label{108}
+u^{2}_{0q}\left[\cos\frac{2z\omega_q}{s} +\frac{m(\Omega^2-\omega^{2}_{q})}{\eta\omega_q}\sin\frac{2|z|\omega_q}{s}\right]\Bigg\}.
\end{eqnarray}
The first term in (\ref{108}) is particle independent and relates to the string only. This part is logarithmically divergent at large momenta. The 
second two terms are related to the particle. For the ground state $\langle\eta^{2}_{q}\rangle=\hbar L/(2\rho\omega_q)$ and at large distances 
$\eta s/(m\Omega^2)\ll z$ 
\begin{equation}
\label{109}
\langle R^2(z)\rangle=\langle R^2(z)\rangle_0+\frac{\hbar}{2\pi m\Omega^2}\frac{s}{|z|}\,,
\end{equation}
where the first term is particle independent and originates from the first term in (\ref{108})
\begin{equation}
\label{110}
\langle R^2(z)\rangle_0=\frac{2\hbar}{\pi\eta}\int^{q_{max}}_{0}\frac{dq}{q}\sin^2qz\simeq\frac{2\hbar}{\pi\eta}\ln zq_{max}\,. 
\end{equation}
We include into (\ref{110}) also the effect of the $s$-mode which doubles the result. 

One can see from (\ref{109}) that the particle, attached to the string, results in its displacements localized along the string. This reminds a
polaronic state of the electron in solids \cite{KITT}.
\subsection{Dissipative motion versus polaronic states}
Usually the electron interaction with photons (string waves in our case) results in a finite width of upper states and therefore in the photon 
emission. This is the dissipative motion of the particle since emitted photons propagate to the infinity carrying away the energy. The corresponding
state is nonstationary which in the classical limit is referred to the electron dissipative motion (\ref{103}), (\ref{104}). 

In this paper a different scenario (polaronic states) is studied. The particle energy spectrum (\ref{13a}) is continuous as in Fig.~\ref{fig2}(b). 
$E_p$ is an exact energy of particle-string states with zero imaginary part. So these states are not decaying that is the particle cannot emit 
photon (propagating sound wave along the string) as in Fig.~\ref{fig2}(a). The ground state energy in Fig.~\ref{fig2}(b) is determined by 
Eq.~(\ref{13b}). 

The polaronic state can arise from the dissipative state. This can happen when the wave, emitted by the particle, is reflected from the end of 
the string and returns. But it takes the infinite time when $L\rightarrow\infty$. 

Another way to form the polaronic state is to act by a non-stationary pulse on the dissipative state. The pulse has to be formed in space to 
reflect emitted photons which, being returned, participate in formation of the polaronic state.
\section{Discussion}
\label{disc}
One can qualitatively explain why photons (string sound waves) are not emitted by polaronic states. This is due to the ``hard'' connection to the 
string. Emission of waves would result in oscillations of the string including the point of particle attachment. This increases the particle
kinetic energy preventing it to lose its total energy and therefore resulting in non-decaying states. These general arguments show that polaronic 
states exist not only in a harmonic potential well.

The issue is that nature allows the continuous non-decaying energy spectrum in a potential well. This conclusion is based on the exact solution of 
the particle-string problem obtained in this paper. There is no contradiction to quantum mechanics since the particle is coupled to the reservoir.

In quantum electrodynamics radiative corrections are small \cite{LANDAU3} and one can say that the electron is not connected ``hard'' to 
electromagnetic coordinates as in the string case. Therefore shifts of discrete energy levels of the electron (the Lamb shift \cite{LANDAU3}) are 
small. This is accompanied by a small broadening of higher levels resulting in photons emission. The Lamb shift occurs since, due to the photon 
influence, the electron ``vibrates'' within the narrow region of $10^{-11}$cm \cite{MIGDAL,BOY}. In that way it probes various parts of the potential 
and therefore slightly changes its energy. Thus in quantum electrodynamics polaronic states are impossible at first sight. 

However, there are indications that states of the polaronic type may exist in quantum electrodynamics. The electron wave function in quantum 
mechanics can have a formal singularity on some line. Such state is not physical. But under electron ``vibrations'' the singularity is smeared out 
within the above narrow region. This anomalous electron-photon state (of the polaronic type), if it exists, is physical. 

That narrow region would play a role of a point where the electron is connected ``hard'' to electromagnetic coordinates and is dragged by them. 
One can treat the electron to be localized in that region. Under the photon emission the narrow region oscillates increasing the electron kinetic 
energy. As in the string case this would prevent the electron to lose its total energy resulting in non-decaying states.

Therefore in quantum electrodynamics one can put a question on possibility of non-decaying states of the electron in a potential well. These states 
are expected of continuous energy spectrum since a singular wave function in quantum mechanics does not lead to levels quantization.

In reality in such non-decaying spectrum the energy can get a small imaginary part because of some non-photon mechanism. In this case we deal with
completely unexpected long-living states of electrons with the continuous energy spectrum. These states can be a basis for laser radiation. Such
radiation is expected to be unusual since it does not correspond to a fixed transition frequency.

Anomalous electron-photon states are supposed to be formed by smearing out of singular wave functions obtained in pure quantum mechanical approach. 
For example, the solution of Schr\"{o}dinger equation $\ln(x^2+y^2)$, singular along the $z$ axis, becomes physical after cutting off the logarithm. 
This occurs on the distance of $10^{-11}$cm by the electron-photon interaction. Formation of such state from usual ones requires an action of the 
certain perturbation of the above spatial range to provide a not small matrix element between the states. For example, it can be irradiation of the 
sample by He ions of the keV energy. Their de Broglie wave length is of that order of magnitude. In other words, it is impossible to arbitrary play 
with choosing of singular solutions of Schr\"{o}dinger equation. 
\section{Conclusions}
The motion of a particle in the potential well is studied when the particle is attached to the infinite elastic string. This is generic with the
problem of dissipative quantum mechanics studied by Caldeira and Leggett \cite{LEGG}. Besides the dissipative motion there is another scenario of 
interaction of the string with the particle attached. Stationary particle-string states exist with string deformations accompanying the particle.
This is like polaronic states in solids. Our polaronic states in the well are non-decaying and with continuous energy spectrum. 

Perhaps these states have a link to quantum electrodynamics. Quantum mechanical wave function, singular on some line, is smeared out by electron 
``vibrations'' due to the interaction with photons. In those anomalous state the smeared singularity position would be analogous to the place where 
the particle is attached to the string. The hypothesis is that, as in the string case, non-decaying electron states in a well with continuous 
energy spectrum can exist.

\acknowledgments
This work was supported by CONACYT through grant number 237439.

\appendix*
\section{Reduction to independent oscillators}
\label{A}
We introduce new variables $\xi_n=r_n$ at $n=1,2,...$ and  $\xi_0=Lx/\sqrt{2}$. The form (\ref{7}) now takes the form
\begin{eqnarray}
\nonumber
&&\frac{2L}{\rho}E=\sum^{\infty}_{n=1}\omega^{2}_{n}\xi^{2}_{n}+\frac{2m\Omega^2}{\rho L}\xi^{2}_{0}+
\sum^{\infty}_{nm=1}(2+\delta_{nm})\dot{\xi}_{n}\dot{\xi}_{m}\\
\label{A1}
&&-4\dot{\xi}_{0}\sum^{\infty}_{n=1}\dot{\xi}_n+2\left(1+\frac{m}{\rho L}\right)\dot{\xi}^{2}_{0}.
\end{eqnarray}
One can apply the linear transformation
\begin{equation}
\label{A2}
\xi_n=\sum_pu_{np}\eta_p,
\end{equation}
where $\eta_p$ is the new variable and matrix elements $u_{np}$ are to be determined to get the form (\ref{A1}) quadratic. The energy (\ref{A1})
looks as 
\begin{equation}
\label{A3}
\frac{2L}{\rho}E=\sum_{pq}\left(A_{pq}\dot{\eta}_{p}\dot{\eta}_{q}+B_{pq}\eta_{p}\eta_{q}\right),
\end{equation}
where 
\begin{eqnarray}
\nonumber
A_{pq}=\sum^{\infty}_{nm=1}(2+\delta_{nm})u_{np}u_{mq}-2\sum^{\infty}_{n=1}(u_{0q}u_{np}+u_{0p}u_{nq})\\
\label{A4}
+2\left(1+\frac{m}{\rho L}\right)u_{0p}u_{0q}\,,\hspace{0.5cm}B_{pq}=\sum^{\infty}_{n=0}u_{np}\kappa_n u_{nq}\,.~~~~
\end{eqnarray}
Here $\kappa_n=\omega^{2}_{n}$ at $n=1,2,...$ and $\kappa_0=2m\Omega^2/(\rho L)$. 

One should choose the matrix $u_{nq}$ in a way to get the relations
\begin{equation}
\label{A5}
A_{pq}=\delta_{pq}\,,\hspace{0.5cm}B_{pq}=\omega^{2}_{q}\delta_{pq}\,,
\end{equation}
where $\omega_{q}$ is the function to be determined. Using Eqs.~(\ref{A2}) and (\ref{A5}) one can easily obtain 
\begin{equation}
\label{A6}
\eta_q=\sum^{\infty}_{n=0}\frac{\kappa_n}{\omega^{2}_{q}}\,u_{nq}\xi_n\,.
\end{equation}
In the equation $B_{pq}-\omega^{2}_{q}A_{pq}=0$, which follows from (\ref{A5}), one can equalize the coefficients at $u_{np}$ to zero. This results
in relations
\begin{equation}
\label{A7}
2u_{0q}=\sum^{\infty}_{m=1}(2+\delta_{nm})u_{mq}-\frac{\kappa_n}{\omega^{2}_{q}}u_{nq},\hspace{0.3cm}n=1,2,...
\end{equation}
and
\begin{equation}
\label{A8}
2u_{0q}\left(1+\frac{m}{\rho L}-\frac{\kappa_0}{2\omega^{2}_{q}}\right)=2\sum^{\infty}_{m=1}u_{mq},\hspace{0.3cm}n=0.
\end{equation}
Excluding $2u_{0q}$ from (\ref{A7}) and (\ref{A8}) we obtain
\begin{equation}
\label{A9}
\left(1+\frac{\rho L}{m}\frac{\omega^{2}_{q}}{\omega^{2}_{q}-\Omega^2}\right)u_{nq}=
\frac{2\omega^{2}_{q}}{\omega^{2}_{n}-\omega^{2}_{q}}\sum^{\infty}_{m=1}u_{mq}\,.
\end{equation}
One can make the summation on $n$ in the both parts of (\ref{A9}) and to cancel $\sum u_{mq}$. Using the formula 
\begin{equation}
\label{A10}
\sum^{\infty}_{n=1}\frac{\cos n\lambda}{n^2-a^2}=\frac{1}{2a^2}-\frac{\pi}{2a}(\sin a|\lambda|+\cos a\lambda\cot\pi a)
\end{equation}
we obtain the relation (\ref{10}) for the function $\omega_q$. The total energy gets the form (\ref{9}).

As follows from Eqs.~(\ref{A8}) and (\ref{A9}), 
\begin{equation}
\label{A11}
u_{nq}=\frac{2m}{\rho L}\,\frac{\omega^{2}_{q}-\Omega^2}{\omega^{2}_{n}-\omega^{2}_{q}}\,u_{0q}.
\end{equation}
Substituting these forms into Eqs.~(\ref{A4}), it is not difficult to check that $B_{pq}=0$ when $p\neq q$. To get the normalization condition
$B_{pp}=\omega^{2}_{p}$ (\ref{A5}) one should choose the proper function $u_{0q}$. One can check that
\begin{equation}
\label{A12}
u_{0q}=\frac{\eta\omega_q}{\sqrt{(m\omega^{2}_{q}-m\Omega^2)^2+\eta^2\omega^{2}_{q}}}\,.
\end{equation}
We omit simple calculations.

\end{document}